\documentclass[12pt]{article}%
\usepackage{amsmath,amsthm,enumerate,graphicx,microtype,amssymb,setspace,verbatim,multirow}
\usepackage[margin= 1in]{geometry}
\usepackage[font={footnotesize}]{caption,subfig}
\usepackage{caption}
\usepackage{setspace}
\usepackage{natbib}
\usepackage{subfig}
\usepackage{amsmath}
\usepackage{amsfonts}
\usepackage{amssymb}
\usepackage{graphicx}
\usepackage{epstopdf}
\usepackage{dsfont}
\usepackage{bbm}
\usepackage{setspace}
\usepackage{soul}
\usepackage{color}
\usepackage{chngcntr}
\usepackage{todonotes}%
\providecommand{\U}[1]{\protect\rule{.1in}{.1in}}
\providecommand{\U}[1]{\protect\rule{.1in}{.1in}}
\newcommand{\distas}[1]{\mathbin{\overset{#1}{\kern\z@\sim}}}
\newsavebox{\mybox}\newsavebox{\mysim}
\newcommand{\distras}[1]{	\savebox{\mybox}{\hbox{\kern3pt$\scriptstyle#1$\kern3pt}}	\savebox{\mysim}{\hbox{$\sim$}}	\mathbin{\overset{#1}{\kern\z@\resizebox{\wd\mybox}{\ht\mysim}{$\sim$}}}}
\makeatother

\begin{document}

\title{\textbf{ Learning Optimal Behavior Through Reasoning and Experiences}%
\textsc{\thanks{We thank Sam Gershman for extensive and helpful discussions. Email addresses:\ Ilut cosmin.ilut@duke.edu, Valchev
valchev@bc.edu. }}}
\author{{Cosmin Ilut}\\\smallskip{Duke University \&\ NBER}
\and {Rosen Valchev}\\\smallskip{Boston College \&\ NBER}}
\date{March 20, 2024}
\maketitle

\begin{abstract}
	
We develop a novel framework of bounded rationality under cognitive frictions that studies learning over optimal behavior through \emph{both} deliberative reasoning \emph{and} accumulated experiences. Using both types of information, agents engage in Bayesian non-parametric estimation of the unknown action value function. Reasoning signals are produced internally through mental deliberation, subject to a cognitive cost. Experience signals are the observed utility outcomes at previous actions. Agents' subjective estimation uncertainty, which evolves through information accumulation, modulates the two modes of learning in a state- and history-dependent way. We discuss how the model draws on and bridges conceptual, methodological and empirical insights from both economics and the cognitive sciences literature on reinforcement learning. 
	
\bigskip


\end{abstract}

\thispagestyle{empty}

\pagebreak\clearpage
\pagenumbering{arabic} \onehalfspacing

\section{Introduction}

There is a deep, interdisciplinary interest in understanding and modeling
cognitive limitations in decision making. Across both economics and cognitive sciences, the literature recognizes two
broad ways in which people learn about and deduce what is the
best course of action in a given situation. The first is \emph{cognition and reasoning}: through
introspective, abstract deliberations, humans can get a better grasp of what is their optimal action in the situation at hand. The second is \emph{accumulated
	experiences}: by observing realized outcomes of past decisions, agents update
their views on the respective benefits of taking those actions in these circumstances. These two sources of
information are conceptually distinct and are both limited: experiences are
observed only along the realized path of situations the agent actually faced in the past, and while abstract
thinking could help the agent deduce outcomes in counterfactual actions and situations, such deliberations are cognitively costly. 

This paper develops a novel framework of constrained-optimal behavior under
cognitive frictions that studies \emph{jointly} learning through reasoning
\emph{and } accumulated experiences. To do so, the paper draws on conceptual,
methodological and empirical insights from both economics and the cognitive
sciences literature on reinforcement learning (RL) (\cite{kaelbling1996reinforcement}, \cite{sutton2018reinforcement}).

In our framework, agents are uncertain about optimal behavior in the
sense of facing subjective uncertainty over the optimal policy and value
functions that characterize their decision problem. In particular, we follow the RL approach of letting agents learn about the \emph{action} value function $Q_{\pi}(a,s)$, which gives the
expected discounted sum of utility when taking action $a$ in state $s$ and following a given policy $\pi(a|s)$ thereafter.\footnote{In the language of RL, this ``action'' value function is closely related and derived from  the ``state'' value function $V(s)$ which is perhaps more familiar to economists.} While the standard approach in economics is to endow the agents with perfect knowledge about the optimal policy function $\pi^{*}(a|s)$ and therefore the action value function $Q_{\pi^{*}}(a,s)$, in our model agents perceive $Q_{\pi^{*}}(a,s)$ as uncertain ex-ante, and gradually learn about it over time. 

Learning can occur
through cognition, which is costly, but beneficial in reducing agents'
uncertainty about the best course of action. Agents trade off that benefit and
cost of engaging cognitive resources, giving rise to a state- and history-
dependent choice of reasoning and thus exhibit constrained-optimal, or
``resource-rational'' behavior. Agents also update beliefs about optimal
behavior based on the experienced flow utility each period. Critically, the
effective precision of \emph{both} reasoning and experiences in informing
behavior is \emph{endogenous}, as a function of the agent's beginning of
period prior beliefs and uncertainty which evolve dynamically and endogenously.

\medskip
\noindent\textbf{Elements of framework.} There are four key specific features of our learning framework. 

First, we model agents' beliefs over the unknown function $Q_{\pi^{\ast}}(a,s)$ as a Gaussian Process (GP) distribution.\footnote{The GP
	distribution can be derived from first principles as the limiting case of
	Bayesian Gaussian kernel regression (eg. \cite{rasmussen2006gaussian},
	\cite{liu2011kernel}).} On the one hand, the GP distribution is methodologically appealing as a prior over the space of functions, as it is very flexible, and also tractable in recursively characterizing the conditional moments of the unknown function.\footnote{In economics, these properties have also recently
	gathered attention to GP, including work like \cite{callander2011searching},
	\cite{bardhi2017optimal}, \cite{dew2019directed},
	\cite{IlutValchevVincent2017} and \cite{ilut2023economic}.}   On the other hand, cognitive sciences increasingly emphasizes a
Bayesian approach and in particular the appeal of GP distributions, both for conceptual and descriptive reasons.\footnote{See for example \cite{griffiths2008bayesian}, \cite{griffiths2010probabilistic} and \cite{gershman2015computational}.}

Second, \emph{given} an estimate of the action value function $\widehat Q_t(a,s)$, agents take constrained optimal actions trading off \textit{exploitation} and \textit{experimentation} incentives. On the one hand, the agent has incentives to choose the action with the highest estimated value $\widehat Q_t( a,s_t)$ at the current state $s_t$ (exploitation). On the other hand, the agent recognizes that $\widehat Q_t(a,s_t)$ is just an uncertain estimate so there is a benefit to exploration and learning more about the unknown $Q_{\pi^*}(a,s)$. We model this desire for experimentation following the concept of maximum entropy reinforcement learning, which essentially the entropy of the action distribution to be bigger than some minimum threshold, thus
ensuring randomization.\footnote{See for example \cite{mnih2016asynchronous},
	\cite{haarnoja2017reinforcement}, \cite{eysenbach2019if}).} In contrast to the standard approach that sets this minimum threshold as an exogenous time-invariant parameter, we let the entropy lower bound be proportional to the remaining \emph{subjective uncertainty} over $Q_{\pi^{*}}(a,s_{t})$. This captures the intuition that experimentation is valuable only to the extent that the $Q$-function is uncertain. Putting it all together, the resulting optimal action policy function takes the form of the \emph{softmax} function widely used in statistics and machine learning (see eg. \cite{sutton2018reinforcement}). 

Turning to the learning modes and the dynamic evolution of the estimates of the action value function, our third modeling feature is learning from experienced utility. Because it is derived from actual observed outcomes, this learning follows the so-called \emph{``model-free''} or ``Q-learning'' dynamic programming solution techniques used in machine learning and cognitive sciences (e.g. \cite{dearden1998bayesian}). In particular,  at the beginning of the period, the agent observes the realization of the new state $s_t$ and also the realized utility $u(s_{t-1}, a_{t-1})$ that she experienced based on last period's choice of action $a_{t-1}$. Knowing the flow utility yesterday and the realized state today implies an \emph{experience-based} informative signal that updates beliefs about $Q_{\pi^*}(a,s)$. The key intuition behind this update, is that the flow utility $u(a_{t-1}, s_{t-1})$ reveals the ``temporal difference'' in the Q-function, i.e. $u(a_{t-1}, s_{t-1}) = Q_{\pi^*}(a_{t-1}, s_{t-1}) - \beta \mathbb{E}_{t-1}(Q_{\pi^*}(a_t, s_t))$. 

Finally, but importantly, within the same objective function that determines optimal actions, we model the benefit and cost of abstract reasoning. By \emph{reasoning} we mean the internal deliberation process through which the agent produces information and learns about $Q_{\pi^{\ast}}(a,s)$ generally. Intuitively, this is akin to an economist trying to solve for the value function globally. As such, our terminology of reasoning is similar to the notion \emph{model-based learning} in the RL literature (\cite{sutton2018reinforcement}). The benefit of reasoning works through the resulting reduction of uncertainty over $Q_{\pi^{*}}(a,s_{t})$. We allow this reduction to impact the objective function in two ways. One is direct utility cost of higher uncertainty (like a cognitive dissonance cost, eg. \cite{aronson1969theory}). 

The second emerges indirectly and endogenously, through an \emph{interaction} between actions and reasoning. In particular, since reasoning decreases current conditional uncertainty, it also weakens the incentive to experiment, by formally lowering the threshold on the action distribution entropy. Intuitively, the agent values the reduction in uncertainty from reasoning because it allows her to select
an action closer to the one with the currently highest value estimate, not having to worry about experimentation. In turn, the cost of reasoning is also in utility terms, quantified as the reduction in entropy achieved by acquiring these deliberation signals, as in information theory (e.g. \cite{Sims2003}), capturing effort as in the cognitive control literature (e.g. \cite{botvinick2004conflict}, \cite{kool2017cost}). 

We show how both reasoning and experience signals can be incorporated in revising beliefs over the unknown $Q$-function using formal non-parametric Bayesian
updating formulas.\footnote{These updates resemble temporal difference
	solution techniques, used extensively in RL (\cite{sutton1988learning}), and in particular their Bayesian, GP-based version, connecting to the approach in
	\cite{engel2003bayes,engel2005reinforcement}.} Put together, these two sources of information update agents' conditional mean and uncertainty over the whole function $Q_{\pi^{\ast}}(a,s)$, providing a recursive structure to the GP distribution entering next period.

\medskip
\noindent\textbf{Key contributions}. Our framework connects to several literature strands.  

In terms of the economics literature,  we use insights from the RL
literature to integrate learning about optimal behavior from experiences \emph{into} models of \emph{resource-rationality} that emphasize a \emph{cost-benefit tradeoff of reasoning}. This tradeoff is shared with various such approaches in economics (including \cite{ilut2023economic}, \cite{sims2003implications},
\cite{gabaix2014sparsity}, \cite{woodford2020modeling}, \cite{alaoui2015cost}
and others) and cognitive sciences (e.g. \cite{gershman2015computational},
\cite{griffiths2015rational}, \cite{shenhav2017toward}). 
In turn, we connect to a growing interest and evidence in
economics on \emph{experience-based learning}, as in
\cite{malmendier2016learning}, \cite{malmendier2021experience}. We differ from this latter approach by studying experiences jointly with
reasoning, as well as by modeling experiences as informative not about the
state law of motion, but instead about the perceived value of taking specific
actions, as in RL.

Our framework innovates within RL modeling itself as well, by building a \emph{dynamic}
cognitive model with several key features emerging endogenously, as follows. \footnote{In this sense, we also differ from recent work by \cite{barberis2023model}, whose main objective is to show how to import
	typical RL frameworks into a specific economic environment - there of asset pricing.} 

First, we obtain and characterize an \emph{endogenous arbitration} between learning from reasoning and
experiences. We use the terminology of ``arbitration'' as in RL, meaning the characterization
of the different weights put on the two modes of learning (``model-based'' vs ``model-free'' learning, in the language of RL). Our framework produces an internally consistent, endogenous arbitration in the form of the state- and history-dependent Bayesian updating weights our agents put on their reasoning and experience signals. The
property that arbitration is based on uncertainty is shared with cognitive and neuroscience work arguing that the brain puts more weight on the learning mode that it deems more reliable (as in \cite{daw2005uncertainty}, \cite{lee2014neural}). Our key contribution here is to incorporate everything in a unified Bayesian framework with  GP priors, which allows (i) prior uncertainty to account for correlation between $Q$-values at
different state/action pairs
, (ii) non-parametric cognitive learning, and (iii) delivers a tractable and parsimonious unified
conditional belief process, as opposed to having to keep track and arbitrarily tie together two separate model-based and model-free estimates.

Second, the framework is characterized by an \emph{endogenous decision to engage reasoning}. The model lets agents treat cognition as an accessible, but costly resource. Thus, it is not
just that, given some model-based and model-free estimates, there is
endogenous arbitration, but the intensity of the model-based
learning and the associated precision of its reasoning signal is time-varying, conditional on states and choices. This implication of adjusting cognition in a state-dependent way is desirable for both empirical and conceptual reasons.
Empirically, it is consistent with evidence and a general desired approach
(see eg. \cite{kool2017cost}) in the neuroscience and RL literatures
emphasizing the cognitive cost of mental simulations versus their benefit. The
latter is here modulated through reduction of uncertainty, where the value of
reducing that uncertainty is (partly) state-dependent. In addition, when the agent values the reduction in uncertainty due to a direct, cognitive dissonance utility cost, this latter mechanism can capture evidence that active deliberation is indeed engaged only when there is sufficient ``conflict'' or uncertainty in prior beliefs (see eg.
\cite{thompson2011intuition}).

Third, the framework describes jointly the evolution of beliefs and the constrained optimal policy function which agents follow in selecting their actions. We endogenously derive a constrained optimal action policy that takes the form of a \emph{softmax} function, which is otherwise exogenously postulated in RL. A key input in that function is the `\emph{temperature}' parameter, which the existing literate treats as an exogenous parameter  -- the lower the temperature, the more the action policy leans away from ``exploration'' and exploits the action with the current highest estimated payoff.  In our model, the temperature parameter is endogenous and is state- and history-dependent, since it depends on the conditional subjective uncertainty the agent still perceives in his estimate of $Q_{\pi^*}(a,s)$. As such, this ``temperature'' of the softmax action policy endogenously evolves as the agent learns. The study of exogenous adjustments of this parameter along the sample path 
has been of large interest in RL, machine learning and constrained
optimization literatures.\footnote{Typical approaches use pre-designed
	strategies tune temperature manually (eg.
	\cite{kirkpatrick1983optimization}, \cite{ackley1985learning}) or adaptively, in deep learning models (eg. \cite{lin2018learning},
	\cite{wang2020meta}).} In that literature, a robustly successful time-varying
temperature (i.e. maximizing reward along simulated paths) is found to be
decreasing through time, since later along the path there is typically less
reason to explore (\cite{sutton2018reinforcement}). Our framework delivers endogenously that result, as uncertainty over $Q_{\pi^{*}}(a,s_{t})$ estimates
naturally declines over time, as well as adds novel state-dependencies in that
logic. \smallskip

Overall, the framework thus results in a dual-learning system, with two components akin to the ``model-based'' and the ``model-free'' modes of learning discussed in the RL literature. Crucially, the paper also extends this literature in providing a unified, formal way of modulating these two modes of learning via
subjective uncertainty, which evolves through information accumulation. Thus, the paper extends both (i) the economics literature, by providing a conceptually new bounded rationality model deeply rooted in cognitive science insights and empirical results, and (ii) the RL literature, by bringing in constrained-optimal maximization approaches and tools typical to information economics.

\section{Framework} \label{sec:framework}

We aim to model the process of economic agents figuring out optimal behavior
using both their abstract thinking abilities, i.e. cognition, and their
accumulated experience with past actions. 

\subsection{A generic recursive problem} To fix ideas, consider a generic recursive problem with
discrete time indexed by $t$, where an agent chooses an action $a_{t}$
(potentially a vector of actions) and is perfectly aware of all
payoff-relevant details of the environment. The problem can be expressed as a
Bellman equation
\begin{equation}
V^{\ast}(s_{t})=\max_{a_{t}\in B(s_{t})}u(a_{t},s_{t})+\beta\mathbb{E}%
_{t}V^{\ast}(s_{t+1}),\label{DP}%
\end{equation}
where $s_{t}$ collects all the relevant state variables, both exogenous and
endogenous. The state follows the known law of motion $F(s_{t+1}|s_{t},a_{t}%
)$, giving the conditional expectation $\mathbb{E}_{t}$ in (\ref{DP}).

The primitives of the environment are the per-period utility function
$u(a_{t},s_{t})$, the time discount factor $\beta$, the budget $B(s_{t})$
defining the set of currently feasible actions $a_{t},$ and the law of motion
of the state $F$. The value function $V^{\ast}(s)$ encodes the continuation
value attached to starting in any particular realization of $s$, when
following the optimal policy function $\pi^{\ast}(a|s)$. The latter is the
optimally selected mapping from any given state $s$ to probabilities over
feasible actions $a$ from equation (\ref{DP}).\footnote{Stating policies in terms of probabilities of
taking actions is useful in developing the framework later. Under full
information, the optimal $\pi^{*}(a|s)$ is naturally degenerate, with
probability $\pi^{\ast}(a_{t}^{*}|s_{t})=1$ for $a_{t}^{*}=\arg\max_{a_{t}\in
B(s_{t})}u(a_{t},s_{t})+\beta\mathbb{E}_{t}V^{\ast}(s_{t+1})$, and zero for
all other actions $a_{t}\neq a_{t}^{*}$.}


Such dynamic problems are at the core of modern economics. And as economists we understand that since dynamic problems
require evaluating whole infinite paths of actions, or in recursive terms, an action
plan with all possible future contingencies, characterizing $V^{\ast}(s)$ is
generally a challenging functional problem (see e.g., \cite{judd1998numerical}
or \cite{bertsekas2019reinforcement}). This is often not fully tractable even
to highly trained economists, and state-of-the-art approximate solution techniques require a lot of sophistication and effort to implement. 

At the same time, in standard economic models is taken as given that the agents themselves always know the optimal policy $\pi^*(a|s)$ and value $V^*(s)$ functions. The difficulty economists face in actually computing the optimal objects $\pi^*(a|s)$ and $V^*(s)$ is  simply abstracted away. We aim to address this apparent paradox, by developing a framework which  puts the
economic agents on a similar footing as economists, by requiring agents to
invest cognitive effort in figuring out $V^{\ast}(s)$, and thus $\pi^{\ast
}(a|s)$ and modeling their gradual learning process. 

In particular our starting point is that in real life people would typically
have \emph{two} sources of information about optimal behavior. The first
source is \emph{experiences}: by observing the per-period utility outcomes of
different actions taken at various states in the past, the agent learns about
$V^{\ast}(s)$. The second source is \emph{cognition and reasoning}: a unique
human characteristic is the ability to think abstractly about the problem at
hand. Through such internal deliberations, agents can learn about the implied
value of taking different courses of action -- for example, this could take
the form of mentally simulating paths forward of possible behavior and
comparing expected utility. These two sources of information are conceptually
distinct and are both limited: experiences are realized only along the
\emph{actual path} taken by the agent, while abstract thinking is a scarce
cognitive resource.


To model learning from both mental simulation and actual experiences, we
connect to the large \emph{reinforcement learning} (RL) literature
(\cite{kaelbling1996reinforcement}, \cite{sutton2018reinforcement}). There an
agent interacting with a dynamic and stochastic environment learns an optimal
control policy for a sequential decision problem, typically a Markov Decision Process.

Using the notation of equation (\ref{DP}), the key object we focus on, as in
RL, is the action-value function $Q_{\pi}(a,s)$. This function gives the
expected utility when taking action $a$ in state $s$ and following a given
policy $\pi(a|s)$ thereafter (not necessarily the optimal $\pi^{\ast}(a|s))$.
The agent wants to know the optimal $Q_{\pi^{\ast}}(a,s)$, defined as
\begin{equation}
Q_{\pi^{\ast}}(a_{t},s_{t})=u(a_{t},s_{t})+\beta E_{t}Q_{\pi^{\ast}}(\pi
^{\ast}(a_{t+1}|s_{t+1}),s_{t+1})\label{qstar}%
\end{equation}
where $\pi^{\ast}(a|s)$ satisfies the Bellman equation (\ref{DP}). The
functions $V^{\ast}(s)$ and $Q_{\pi^{\ast}}(a,s)$ are implicitly related, as
$V^{\ast}(s_{t})=\max_{a_{t}\in B(s_{t})}Q_{\pi^{\ast}}(a_{t},s_{t})$, but as we detail later, it will be useful to work with $Q_{\pi^{\ast}}(a_{t},s_{t})$.

\subsection{Human learning with Gaussian Processes\label{learnGP}}

We model agents' beliefs over the unknown function $Q_{\pi^{\ast}}(a,s)$ as
Gaussian Process (GP) distributions. In particular, at the beginning of time
agents have the initial prior
\begin{equation}
Q_{\pi^{\ast}}(a,s)\sim GP(\widehat{Q}_{0}(a,s),{\widehat{\Sigma}}%
_{0}(a,s,a^{\prime},s^{\prime}))\label{prior}%
\end{equation}
where the mean function is $\widehat{Q}_{0}(a,s)=\mathbb{E}(Q_{\pi^{\ast}%
}(a,s))$ and the variance-covariance function is $\widehat{\Sigma}%
_{0}(a,s,a^{\prime},s^{\prime})=Cov(Q_{\pi^{\ast}}(a,s),Q_{\pi^{\ast}%
}(a^{\prime},s^{\prime}))$. The defining feature of a Gaussian Process distribution is that for any two action-state pairs $(a,s)$ and $(a',s')$, the values $Q_{\pi^*}(a,s)$ and  $Q_{\pi^*}(a',s')$ have a joint-Normal distribution with mean and variance-covariance given by $\widehat Q_0$   and $\widehat \Sigma_0$.

\textbf{Why learning with GP?} There is a variety of motivating insights and arguments to using GP distributions and in particular so in the context of modeling human cognition. 

On the one hand, GP distributions are methodologically appealing due to their
flexibility and tractability. In particular, GP distributions extend the
familiar Kalman filter to learning about functions and the law of motion for
the conditional mean and variance can be easily characterized recursively. As
a result, the \emph{conditional} beliefs in \emph{each} period also follow a
GP distribution, with properly updated mean and variance functions, as
detailed below. In this formulation the resulting variance-covariance
function $\widehat{\Sigma}_{0}(a,s,a^{\prime},s)$ encodes the agent's prior
view of how likely correlated are the function's values at different points
(here pairs $(a,s)$ and $(a^{\prime},s^{\prime})$), and therefore is inducing
a measure of proximity, or similarity, between those points.

On the other hand, the cognitive sciences literature increasingly emphasizes a
Bayesian approach and in particular the appeal of GP distributions. Indeed, at
a broad level, cognitive sciences emphasize that cognition is fundamentally
related to forming uncertain conjectures from partial or noisy information,
and thus a probabilistic framework is particularly well suited both
conceptually and in terms of accounting for data on observed behavior (eg.
\cite{chater2006probabilistic}, \cite{griffiths2008bayesian},
\cite{griffiths2010probabilistic}).

The GP distribution in particular has been increasingly used in the cognitive
sciences literature, building on connections to statistics and machine
learning (on the latter see eg. \cite{barber2012bayesian}). The reason is in
the recent experimental and neuroscience evidence that the human brain's
learning process is well described by GP (see for example
\cite{gershman2015computational} for a survey and \cite{schulz2018putting},
\cite{wu2018generalization}, \cite{wu2021inference} for evidence using
bandit-like tasks).

Even more specific to RL, the Bayesian approach and the GP distribution have
also found applications in this RL literature (see eg.
\cite{engel2005reinforcement}). As surveyed in \cite{ghavamzadeh2015bayesian},
this approach can provide a tractable and coherent way to model the
exploitation-exploration tradeoff, fundamental to learning from experiences,
as a function of subjective uncertainty over $Q_{\pi^{\ast}}(a,s)$ estimates,
as well as offering a formal way to incorporate prior beliefs into the
action-selection algorithms.

In building our framework, we nevertheless emphasize that these motivating
insights from cognitive science have not been connected in a single framework
that \emph{interacts} learning from abstract thinking and from experiences,
and have also not been previously applied to economic models.

\subsection{Learning from experienced outcomes}
We start by detailing how we formalize the learning through experienced outcomes and its connection to the concept of  ``model-free'' learning in machine learning. We will consider this type of learning as free of cognitive costs, as it accumulates by the agent simply experiencing utility flows of the specific actions she has taken in the past, and not through abstractly thinking about contingencies and counterfactuals. 

Consider a
typical period $t$. The agent enters the period with some prior beliefs,
conditional on the prior information set $\mathcal{I}_{t-1}$. As anticipated
earlier, given the time zero prior in equation (\ref{prior}), the conditional $t-1$
beliefs also have a GP distribution: 
\begin{equation}
	Q_{\pi^{\ast}}(a,s)|\mathcal{I}_{t-1}\sim GP(\widehat{Q}_{t-1}(a,s),{\widehat
		\Sigma}_{t-1}(a,s,a^{\prime},s^{\prime}))\label{qpriors}%
\end{equation}
where $\widehat Q_{t-1}(a,s)= \mathbb{E}( Q_{\pi^{*}}(a,s) | \mathcal{I}%
_{t-1})$ and $\widehat\Sigma_{t-1}(a,s,a^{\prime},s^{\prime}) = Cov(
Q_{\pi^{*}}(a,s), Q_{\pi^{*}}(a^{\prime},s^{\prime}) | \mathcal{I}_{t-1})$ follow a recursive formulation detailed below. 

At the beginning of the period, the agent observes the realized utility $u(s_{t-1}, a_{t-1})$ that she experienced based on last period's choice of action $a_{t-1}$.  In addition, the time $t$ shock also realizes and the agent observes the realization of the new state variable $s_t$.  Knowing the flow utility yesterday and the realized state today implies an \emph{experience-based} informative signal that updates beliefs about $Q_{\pi^*}(a,s)$. This update, \emph{along the realized path} of state action pairs $(a_t, s_t)$, follows the so called ``model-free'' or ``Q-learning'' dynamic programming solution techniques used in machine learning and cognitive sciences (e.g. \cite{dearden1998bayesian}). 

The key intuition behind this update, is that the flow utility $u(a_{t-1}, s_{t-1})$ reveals the ``temporal difference'' in the Q-function, that is from the Bellman equation we can express

\[   u(a_{t-1}, s_{t-1}) = Q_{\pi^*}(a_{t-1}, s_{t-1}) - \beta \mathbb{E}_{t-1}( Q_{\pi^*}( \pi^*(a_{t}|s_{t}), s_{t} ))  \]

Given beliefs $\widehat Q_{t-1}(a,s)$, one can then compute the deviation from the above equation that any specific beliefs $\widehat Q_{t-1}(a,s)$ imply, and adjust them accordingly. To do this directly from the above equation, this requires computing the expectation $\mathbb{E}_{t-1}( Q_{\pi^*}( \pi^*(a_{t}|s_{t}), s_{t} )) $ which integrates over all possible states $s_{t}$ that could have realized at time $t$. This integration is computationally and conceptually a complex step, and thus, typically machine learning applications actually use the more robust approach which utilizes instead the approximation  
\begin{equation} \label{eq: Qlearn}  u(a_{t-1}, s_{t-1}) = Q_{\pi^*}(a_{t-1}, s_{t-1}) - \beta  Q_{\pi^*}( \pi^*(a_{t}|s_{t}), s_{t} ) \end{equation}
where the temporal difference on the right-hand side is not between the   Q-function at $(a_{t-1}, s_{t-1})$ and the average across all possible   states $s_{t}$, but only the difference with the value of the Q-function in the actually realized state $s_{t}$. 

This approach is ``robust'', in the sense that it does not need to compute the expectation $\mathbb{E}_{t-1}$ and also does not need to assume full knowledge of the transition probabilities $F(s_{t+1}| s_t, a_t)$. The agent is simply sitting at time $t$ and observing the actual realization of $s_{t}$, and then only looks back at the old state-action pair $(a_{t-1}, s_{t-1})$ and the realized utility $u(a_{t-1}, s_{t-1})$. Still, using this approximation to update estimates of the $Q$-function is \emph{asymptotically consistent}, in that if agents visit all possible states in the support of $s_t$ with positive probability, then the update based on the approximation will eventually converge to the true $Q_{\pi^*}(a,s)$ as $t\rightarrow \infty$. 

Also worth stressing is that this ``model-free'' update is very simple to do, as it just compares the agent's realized utility yesterday and her current expectation of the $Q$-function at the realized state today. As such, it is straightforward to model this experiential updating as ``cognitively free'' -- as something that just comes by default to the agent. 

Formally, the agent perceives the following experience signal 
\[ \eta_{t}^E \equiv  u(a_{t-1}, s_{t-1}) \approx Q_{\pi^*}(a_{t-1}, s_{t-1}) - \beta  Q_{\pi^*}( \pi^*(a_{t}|s_{t}), s_{t} )  \]

And lastly, since our framework keeps
track of beliefs about $Q_{\pi^{\ast}}(a,s)$, but not explicitly of
beliefs over $\pi^{\ast}(a|s_{t})$, for tractability it is convenient to
approximate the latter by setting $\pi^{\ast}(a|s_{t})=\pi_{t-1}%
^{greedy}(a|s_{t})$.\footnote{This approximation can be in principle relaxed
	by adding a further layer of noise.} Thus, we assume that the agent perceives the following structure of the experience signal 
\[ \eta_{t}^E \equiv  u(a_{t-1}, s_{t-1}) \approx Q_{\pi^*}(a_{t-1}, s_{t-1}) - \beta  Q_{\pi^*}( \pi_{t-1}^{greedy}(a_{t}|s_{t}), s_{t} )  \]

This expression for $\eta_{t}^{E}$ can be readily used to compute a formal Bayesian update based on this experiential information. Specifically, 
\begin{equation}
	\widehat{Q}_{t}^E(a,s) = \widehat Q_{t-1}(a,s) +\alpha_{t}^{E}(a,s)\left[
	\eta_{t}^{E} - \left( \widehat Q_{t-1}(a_{t-1}, s_{t-1}) - \beta \widehat Q_{t-1}( \pi_{t-1}^{greedy}(a_{t}|s_{t}, s_{t})) \right) \right]  \label{qmodelfree}%
\end{equation}
where 	$\widehat{Q}_{t}^E(a,s) \equiv \mathbb{E}( Q_{\pi^*}(a,s) | \mathcal{I}_{t-1}, \eta_t^E)$ is not quite the end-of-period $t$ beliefs, as beliefs will potentially be further updated via the abstract reasoning we describe below. Moreover, the  signal to noise ratio $\alpha_{t}^{E}(a,s)$ can be derived in a straightforward say by evaluating
\begin{equation}
	\alpha_{t}^{E}(a,s)=\frac{Cov(\eta_{t}^{E},Q_{\pi^{\ast}}(a,s)| \mathcal{I}%
		_{t-1})}{Var(\eta_{t}^{E}|\mathcal{I}_{t-1}%
		)}\label{alphaq}%
\end{equation}

The signal also reduced the conditional uncertainty facing the agent, as
\begin{eqnarray*}
	\Sigma_t^E(a,a'; s_t) & \equiv & Cov(
	Q_{\pi^{*}}(a,s_t), Q_{\pi^{*}}(a^{\prime},s_t) | \mathcal{I}_{t-1}, \eta_t^E)\\
	& = & \widehat \Sigma_{t-1}(a,a';s_t) - \alpha_t^E(a,s)Cov(\eta_{t}^{E},Q_{\pi^{\ast}}(a,s)| \mathcal{I}%
	_{t-1})		
\end{eqnarray*}

\subsection{Learning through abstract reasoning} 

After updating beliefs for free, each period $t$ the agent decides how
intensely to engage in cognitively costly abstract reasoning, which produces further information about $Q_{\pi^*}(a,s)$.  And then, after updating beliefs based on all sources for time $t$ information, the agent optimally chooses an action policy $\pi_{t}(a|s_{t})$. To
understand these choices, consider first the reasoning process.

\subsubsection{Reasoning}

By \emph{abstract reasoning} we mean the internal deliberation process through which the agent thinks abstractly about his decision problem, and 
 learns about $Q_{\pi^{\ast}}(a,s)$ generally.\footnote{For ease of exposition, we assume
that the action space is discrete with cardinality $N_{a} = |a|$, but this can
be relaxed and the framework can be defined on a continuous space of actions
as well.} We will remain agnostic about the specific mode of deliberation the agent engages in, but for example they can be trying to do value function iteration or just thinking forward through the decision tree of their problem. This kind of reasoning is deliberate and abstract, as it tries to deduce new information about $Q_{\pi^*}(a,s)$ over and above the simple experience of the flow utility given past choices. It is a potentially powerful source of information for the agent, but is mentally costly. 

Formally, each period the agent can generate a vector of reasoning signals $\eta_{t}^{R}$ as unrestricted linear
functions of $Q_{\pi^{\ast}}(a,s_{t})$
\begin{equation}
\eta_{t}^{R}=\Omega_{t}^{\prime}Q_{\pi^{\ast}}(a,s_{t})+\varepsilon_{\eta
,t}\label{etat}%
\end{equation}
where $\varepsilon_{\eta,t}\sim N(0,\Sigma_{\eta,t})$ and $\Omega_t$ is a $N_a \times N_a$ matrix. The agent optimally
chooses the structure of the matrix $\Omega_{t}$ and the noise variance matrix
$\Sigma_{\eta,t}$, subject to a cost-benefit tradeoff introduced below in subsection \ref{choice}.

The vector of signals $\eta_{t}^{R}$ then updates beliefs over  
$Q_{\pi^{\ast}}(a,s)$ at all pairs $(a,s)$:
\begin{equation}
\widehat{Q}_{t}(a,s)=\widehat{Q}_{t}^{E}(a,s)+\alpha_{t}^{R}(a,s)\left(
\eta_{t}^{R}- \Omega_{t}^{\prime}\widehat{Q}_{t}^E(a,s_{t})\right)
\label{qmodelbased}%
\end{equation}
where the signal-to-noise ratio $\alpha_{t}^{R}(a,s)$ is given by
\begin{equation}
\alpha_{t}^{R}(a,s)= Cov\left(Q_{\pi^{\ast}}(a,s), \Omega_{t}^{\prime}%
Q_{\pi^{\ast}}(a,s_{t})| \mathcal{I}_{t-1} , \eta_t^E \right) \left(Var (\Omega_{t}^{\prime}Q_{\pi^{\ast}}%
(a,s_{t})  | \mathcal{I}_{t-1}, \eta_t^E) + \Sigma_{\eta,t}\right)^{-1} \label{alphar}%
\end{equation}
which can be tractably computed from $\widehat\Sigma_{t-1}(a,a^{\prime
},s,s^{\prime})$ and the update based on $\eta_t^E$ described above. 

Importantly,the reasoning signals further reduce uncertainty, so that
\begin{equation}
\widehat\Sigma_{t}(a,a^{\prime};s_{t})\equiv Cov(
Q_{\pi^{*}}(a,s_t), Q_{\pi^{*}}(a^{\prime},s_t) | \mathcal{I}_{t}) = \Sigma_{t}^E(a,a'; s_t) - \alpha_t^R(a,s)\Sigma_{t}^E(a, \Omega_t a; s_t)' \label{sigma_r}
\end{equation}denote the posterior variance function over the vector

\textbf{Interpretation.} Our terminology of reasoning is similar to the notion of \emph{planning} or \emph{model-based learning} in the computational and RL literature (\cite{sutton2018reinforcement}). Reasoning can in practice work
through various specific processes. Consider for example a specific tool used in the model-based RL -- the simulation of a limited number of future paths of
actions and resulting states and utility flows in the future (see for example
\cite{moerland2023model}). In our framework, this simulation is captured by the
noisy signals $\eta_{t}^{R}$ that update the prior beliefs of the agent. In a related way, by producing new information internally to the agent, without being observed by an outsider, the notion of reasoning here also resembles in economics that of `fact-free learning' in \cite{gilboa2005fact} and
\cite{alaoui2015cost}.

However, our proposed framework purposefully abstracts from the specifics of
the mental method behind $\eta_{t}^{R}$. Instead it aims to capture a key
tradeoff that many different solution methods share, namely that taking
more computational steps in any given method (e.g. simulate more and further
paths forward), leads to (i) higher accuracy of the resulting solution, but
(ii) is cognitively costlier, as we describe next.

Overall, by incorporating both signals $\eta_t^E$ and $\eta_t^R$ in the conditional beliefs of our agent we are integrating, in an endogenous way, both ``model-free'' and ``model-based'' learning, which are the two main types of learning   discussed in RL. 

This joint integration of both types of learning is present in both some recent treatments of RL  (see
\cite{sutton2018reinforcement}) and also  in classical RL algorithms such as
Dyna-Q (\cite{sutton1990integrated,sutton1991dyna}).\footnote{These algorithms
	include many extensions, such as Dyna-Q+ (\cite{sutton1990integrated}), which
	like in the current framework further stimulates exploration.}  Nevertheless, the RL literature usually resorts to ad-hoc assumptions on how the two types of learning are mixed together, while our framework proposes an endogenous arbitration between the two types of signals based on formal Bayesian updating notions. 

Furthermore,
the specific formulation of belief updating for both signals in equations
(\ref{qmodelbased}) and (\ref{qmodelfree}) resembles temporal difference
solution techniques, used extensively in RL (\cite{sutton1988learning}), and
in particular their Bayesian, GP-based version, connecting to the approach in
\cite{engel2003bayes,engel2005reinforcement}.
\subsubsection{Joint choice over reasoning and actions at time $t$\label{choice}} 

Equations (\ref{qmodelbased}) and (\ref{sigma_r}) describe the updated beliefs, taking as given the structure for the signal in equation (\ref{etat}). Critically, given the current state $s_t$ and prior beliefs in equation (\ref{qpriors}), we let the agent jointly choose her optimal reasoning structure (i.e. $\Omega_{t}$ and $\Sigma_{\eta,t}$) and the action policy $\pi_t(a | s_t)$ that he will follow this period.

\bigskip\noindent\textbf{Objective function.} We let the agent maximize the
following joint objective function
\begin{equation}
	\max_{\pi_{t}(a|s_{t}), {\Sigma}_{\eta,t}, \Omega_{t}}\underbrace{\sum
		_{a}\widehat{Q}_{t}(a,s_{t})\pi_{t} (a|s_{t})}%
	_{\substack{\text{exploitation benefit}}}-\underbrace{w\sum_{a}\sigma_{t}%
		^{2}(a,s_{t})}_{\substack{\text{cognitive dissonance cost}}}-\underbrace{\frac{\kappa
		}{2}\ln\left[  \frac{|\Sigma_{t}^E(a,s_{t},a^{\prime},s_{t})|}{|{\Sigma}%
			_{t}(a,a^{\prime};s_{t})|}\right] }_{\substack{\text{reasoning cost}}}
	\label{obj}%
\end{equation}
where recall that $\Sigma_{t}^E(a,s_{t},a^{\prime},s_{t})$ is the 
variance conditional on both the beginning-of-period information set $\mathcal{I}_{t-1}$  and the experience signal $\eta_t^E$, while  $\Sigma_{t} ^{R}(a,a^{\prime};s_{t})$ is the end-of-period posterior variance
defined in equation (\ref{sigma_r}), after also updating with the chosen reasoning signals. Here
\[
\sigma_{t}^{2}(a,s_{t}) \equiv\Sigma_{t}(a,a;s_{t})
\]
denotes the diagonal entries of the posterior variance $\Sigma_{t}(a,a'; s_t)$.

The objective function in (\ref{obj}) is subject to two types of constraints
on the action distribution. First, is that of feasibility: $\pi(a|s_{t}) = 0$
for actions that do not satisfy the budget constraint $a \notin B(s_{t})$, and
that the action distribution probabilities sum to one:
\[
\sum_{a}\pi_{t}(a|s_{t})=1
\]
Second, the action distribution $\pi_{t}(a|s_{t})$ is subject to an entropy
constraint:
\begin{equation}
	\underbrace{-\sum_{a}\ln(\pi_{t}(a|s_{t}))\pi_{t}(a|s_{t})}%
	_{\substack{\text{entropy of action distribution}}}\geq h\sum_{a}{\sigma}_{t}
	^{2}(a,s_{t}) \label{entropyactions}%
\end{equation}

\textbf{Interpretation.} The objective function in (\ref{obj}) and the entropy
constraint in (\ref{entropyactions}) allow our framework to capture jointly a
variety of forces of interest emphasized in the RL and cognitive science
literature, as follows.

The first term in the objective of equation (\ref{obj}) reflects the agent's
\textit{benefit of exploitation}, or ``greediness'' in the RL language.
This force incentivizes the agent to choose the action $a_{t}$ with the
current highest estimated value $\widehat{Q}_{t}(a_{t},s_{t})$, where the
latter is given in equation (\ref{qmodelbased}). The second term captures a
possible \emph{cognitive dissonance cost} (\cite{aronson1969theory},
\cite{akerlof1982economic}). Essentially, when the primitive disutility
parameter $w>0$, there is a disutility cost of uncertainty over the values
associated to each action. This cost generates a benefit of reasoning that
goes purely through the reduction of that dissonance and the disutility of facing uncertainty about the Q-function. This mechanism is qualitatively similar to the one used in \cite{ilut2023economic}, and is distinct from the novel 
exploitation-experimentation tradeoff described below. 

The third term  in the objective function measures the \emph{cognitive cost of reasoning} as proportional to  the information content of signals $\eta_{t}^{R}$, with information flow quantified as the reduction in entropy achieved by acquiring these signals, following a standard information theory approach, e.g. \cite{Sims2003}. The reasoning cost captures the fact that increasing the reasoning intensity (e.g.
mentally simulating more paths forward) requires higher cognitive effort. Like
in the cognitive control literature (e.g. \cite{kool2017cost},
\cite{botvinick2004conflict}, \cite{botvinick2014computational}), the constant marginal cost $\kappa>0$ can be interpreted as the opportunity cost
of cognitive capacity, which can be otherwise employed on other,
outside-the-model tasks.\footnote{Thus, $\kappa$ will be higher if individuals have a
	higher opportunity cost of cognitive capacity or if their particular
	deliberation process behind $\eta^{R}_{t}$ takes longer to produce a given
	amount of information. In addition, $\kappa$ will also be higher if the
	environment is more complex and it is thus objectively harder to come up with
	insights on the unknown optimal $Q$-function.}

In turn, the interpretation and aim of the constraint in (\ref{entropyactions}%
) is to capture a \textit{desire for experimentation}. As in typical bandit
problems, as long as the value function $Q_{\pi^{*}}(a,s)$ is uncertain, there
is a desire to explore the function in other parts of the state space.
Following the RL literature, we parsimoniously model this desire to experiment
using the idea of \textquotedblleft\textit{entropy regularization}%
\textquotedblright, or maximum entropy RL, (eg. \cite{mnih2016asynchronous},
\cite{haarnoja2017reinforcement}, \cite{eysenbach2019if}). This regularization
requires the entropy of the action distribution (the LHS of equation
(\ref{entropyactions})) to be bigger than some minimum threshold, thus
ensuring randomization. With this constraint, the chosen action distribution
$\pi_{t}(a|s_{t})$ is not degenerate, but always puts some probability of
exploring any feasible action, thus capturing the exploration incentive
inherent to dynamic learning problems. Importantly, in our framework this
entropy constraint builds in that experimentation is valuable only to the
extent to which the $Q$-function is uncertain -- indeed, if $Q_{\pi^{*}%
}(a,s_{t})$ was known, there would be no point to experiment as there is
nothing further to learn. Hence, when the experimentation parameter $h>0$, in
the RHS of equation (\ref{entropyactions}) the entropy lower bound is set
\emph{proportional} to the remaining subjective uncertainty over $Q_{\pi^{*}%
}(a,s_{t})$, as measured by $\sum_{a} \sigma_{t}^{2}(a,s_{t})$.

\subsubsection{Optimal policy action\label{actions}}

Let $\delta_{t}$ denote the Lagrange Multiplier on the constraint in equation
(\ref{entropyactions}). Given the set of feasible actions in $B(s_{t})$, the
optimal policy action is
\begin{equation}
	\widehat{\pi}_{t}(a|s_{t})=
	\begin{cases}
		\frac{\exp\left( \frac{\widehat{Q}_{t}(a,s_{t})}{\delta_{t}}\right) }%
		{\sum_{a}\exp\left( \frac{\widehat{Q}_{t}(a,s_{t})}{\delta_{t}}\right) } &
		\text{ if }{\delta}_{t}>0\\
		\widehat{\pi}_{t}^{greedy}(a|s_{t}) & \text{ if }{\delta}_{t}=0
	\end{cases}
	\label{pi}%
\end{equation}
where $\widehat{\pi}_{t}^{greedy}(a|s_{t})$ is the degenerate greedy policy
\begin{equation}
	\widehat{\pi}_{t}^{greedy}(a|s_{t})=
	\begin{cases}
		1, \quad\text{for} \hspace{0.15cm} a=\widetilde{a_{t}}\equiv\arg\max_{a_{t}\in
			B(s_{t})}\widehat{Q}_{t}(a_{t},s_{t})\\
		0 ,\quad\forall\hspace{0.1cm} a\neq\widetilde{a_{t}}%
	\end{cases}
	\label{pi_greedy}%
\end{equation}


\textbf{Softmax.} The optimal policy in (\ref{pi}) takes the form of the
\emph{softmax} function widely used in statistics and machine learning. This
function takes the expected rewards of following any given action and
transforms them into action probabilities (eg. \cite{sutton2018reinforcement}%
). However, in that literature $\delta_{t}$, known as the `\emph{temperature}'
parameter, is typically exogenous. The lower is $\delta_{t}$, the more the
expected rewards affect the probability of taking actions. In our model,
$\delta_{t}$ is \emph{endogenous} and state and time-dependent. Specifically,  it will be high for states $s_t$ where the agent faces a higher uncertainty (and thus $\sum_a \sigma_t^2(a,s_t)$ is high).

In economics, a similar softmax function describes the multinomial logic model
of choices (eg. \cite{luce1959individual}), typically micro-founded from a
random utility model where valuations of actions are subject to additively
separable and independently distributed shocks drawn from the extreme-value
distribution (eg. \cite{mcfadden1974conditional}). Instead, the optimal policy
$\widehat{\pi}_{t}(a|s_{t})$ proposed here does not rely on such random
utility shocks.

A complementary micro-foundation for logit choice in the literature is based
on rational inattention (\cite{sims2003implications}). In that work (eg.
\cite{matejka2014rational} in economics, or the policy compression work of
\cite{lai2021policy} in cognitive sciences), the entropy of the actions
distribution, i.e. LHS of equation (\ref{entropyactions}), speaks to an
information flow constraint across actions, given signals about the rewards
associated to those actions. Here instead the entropy on actions distribution
speaks to the entropy regularization logic in RL, per above.

To see that comparison, consider setting $h=0$ in constraint
(\ref{entropyactions}). In that case, the constraint would not be binding and
the optimal action becomes $\widehat{\pi}_{t}^{greedy}(a|s_{t})$. Critically,
even without a role for the entropy on \emph{actions} distributions, when
$\kappa>0$ there is still an entropy informational cost on \emph{reasoning
	signals} in objective (\ref{obj}). So, when $h=0$ reasoning is here still
\emph{imperfect}, but given imperfect posterior Bayesian estimates
$\widehat{Q}_{t}(a,s_{t})$, the agent would take the greedy perceived action
with probability one. The extreme case of $h=0$ would thus have no role for
exploration, a fundamental feature to dynamic learning problems.\footnote{This
	discussion also implies that the framework could extend the RHS of constraint
	(\ref{entropyactions}) to be $c+h\sum_{a}\widetilde{\sigma}_{t}^{2}(a,s_{t})$.
	Even if uncertainty over $\widehat{Q}_{t}(a,s_{t})$ would disappear, when the
	new parameter $c>0$, it would still difficult for the agent to compare all the
	$\widehat{Q}_{t}(a,s_{t})$ at current state $s_{t}$ for all possible actions,
	generating stochastic choice like in the policy compression logic. That
	integrated softmax could then nest both policy compression and experimentation
	sources.}

At another extreme, consider \emph{the limit of costless reasoning}, i.e.
$\kappa=0.$ In that case, $\sigma_{t}^{2}(a,s_{t})=0$ as cognitive effort is
chosen to be infinitely high, and agents learn $Q_{\pi^{*}}(a,s)$ perfectly.
As there is no remaining uncertainty, the entropy regularization constraint
(\ref{entropyactions}) would not bind, $\delta_{t}=0,$ and the optimal action
follows the greedy policy. The objective in equation (\ref{obj}) would then
recover exactly the problem solved under full rationality (see equations
(\ref{DP}) and (\ref{qstar})) since $\widehat Q_t(a,s) = Q_{\pi^{*}}(a,s).$

	\subsubsection{Optimal reasoning signal structure\label{optimal_reasoning}}
	
	The optimal reasoning choice consists of the agent deciding on the matrix
	$\Omega_{t}$ and signal noise variance matrix $\Sigma_{\eta,t}$. To solve this
	problem, we use insights from rate-distortion theory.
	
	Let $\lambda_{t}^{(i)}$ denote the eigenvalues of posterior variance
	${\widehat\Sigma}_{t}(a,a^{\prime};s_{t})$. Using the standard property
	that a matrix determinant equals the product of its eigenvalues, $\ln(|
	\widehat\Sigma_{t}(a,a^{\prime};s_{t})|) = \sum_{i} \ln(\lambda_{t}%
	^{(i)}),$ it follows that optimal reasoning can be expressed as choosing the
	resulting eigenvalues $\lambda_{t}^{(i)}$ optimally.

 For notational convenience,  it is useful to sort the eigenvalues in descending order so
	\[
	\lambda_{t}^{(1)}\geq\lambda_{t}^{(2)}\geq...\geq\lambda_{t}^{(N)}%
	\]
	By standard properties of variance matrices $\Sigma_{\eta,t}$ is positive
	definite, and thus the (sorted) eigenvalues of the posterior variance
	$\widehat\Sigma_{t}(a,a^{\prime};s_{t})$ must be weakly smaller than those
	of the ``prior'' variance $\widehat\Sigma_{t}^E(a,a^{\prime};s_{t})$, hence
	$\lambda_{t}^{(i)}\leq\lambda_{E,t}^{(i)}$ (where $\lambda_{Et}^{(i)}$ denote the sorted eigenvalues of $\widehat\Sigma_{t}^E(a,a^{\prime};s_{t})$). Taking this constraint into
	account and optimizing the objective function over $\lambda_{t}^{(i)}$, the
	optimal reasoning choice implies
	\begin{equation}
		\lambda_{t}^{(i)}=\min\left\{ \frac{\kappa}{w+\delta_{t}h},\lambda_{E,t}%
		^{(i)}\right\}  \label{optlambda}%
	\end{equation}

	Hence, the agent has an optimal target level for uncertainty and wants to only
	reduce eigenvalues that currently bigger than that threshold $\frac{\kappa
	}{w+\delta_{t}h}$ -- a property known as reverse `water filling' in
	rate-distortion theory (see \cite{cover1999elements}, chapter 13). Thus, we
	can show that the optimal signal structure is for $\Omega_{t}$ to be the
	matrix of eigenvectors of $\widehat\Sigma_{t}^E(a,a^{\prime};s_{t})$, and
	$\Sigma_{\eta,t}$ be a diagonal matrix with entries\footnote{In practice, the
		eigenvalues are likely to decline in  value quickly, so only a few of these
		signals will typically have non-infinite  noise variance making the solution
		very tractable and easy to implement.}
	\begin{equation}
		\sigma_{i,t}^{2}=
		\begin{cases}
			\frac{\kappa\lambda_{E,t}^{(i)}}{\lambda_{E,t}^{(i)}( w + \delta_t h)-\kappa} & \text{when}
			\quad\lambda_{E,t}^{(i)}> \frac{\kappa}{ w + \delta_t h}\\
			\infty & \text{when} \quad\lambda_{E,t}^{(i)}\leq \frac{\kappa}{ w + \delta_t h}
		\end{cases}
	\end{equation}


Put together, equations (\ref{pi}) and (\ref{optlambda}) uncover a tight intuitive
\emph{interaction} between actions and reasoning intensity. For example, take
states $s_{t}$ where uncertainty is high and thus the endogenous $\delta_{t}$
is larger. By (\ref{pi}), a tighter constraint induces the agent to explore
more and thus deviate from the action with the currently highest perceived $\widehat
Q_{t}(a,s_{t})$. On the other hand, reasoning  lowers uncertainty and thus relaxes this
constraint, lowering $\delta_t$ and leads to a policy function $\pi_t(a|s_t)$ that selects actions $a$ with a higher expected payoff $\widehat Q_t(a,s_t)$.  

Looping back to the \emph{choice} of how much cognitive effort to invest in reasoning,  from (\ref{optlambda}) we see that when $\delta_{t}$ is larger, the
agent employs more cognition to achieve a greater reduction in  uncertainty (the optimal target for $\lambda_t^{(i)}$ is lower). Intuitively, she
values that reduction in uncertainty precisely because it allows her to select
an action closer to the currently perceived greedy policy, not having to worry
about experimentation.

Overall, the framework is well suited to apply to a range of concrete economic environments, some of which we actively investigate in our own work in progress. For example, a standard consumption-savings problem in a simple \cite{aiyagari1994uninsured} framework is a transparent and widely-studied setting in economics. Indeed, the consumption-saving decision is a fundamental mechanism in a number of different economic settings, and has been used as a laboratory for other recent bounded rationality and behavioral papers, such as for example \cite{ilut2023economic} and \cite{lian}. Due to its core elements, the bounded rationality framework proposed here appears promising in delivering endogenously empirically documented properties for consumption that are puzzling for standard fully-rational models: (a) experience effects of past income shocks on future consumption choices, (b) high sensitivity of consumption to income shocks; (c) large heterogeneity in consumption responses, through stochastic signals and choice.

\pagebreak\pagestyle{empty}
\bibliographystyle{aer}
\bibliography{complexity_bibl}

\end{document}